\documentclass[10pt,sigconf,letterpaper,authorversion]{acmart}

\usepackage{enumitem}
\usepackage{xurl}
\copyrightyear{2021} 
\acmYear{2021} 
\setcopyright{acmlicensed}\acmConference[ANRW '21]{Applied Networking Research Workshop}{July 24--30, 2021}{Virtual Event, USA}
\acmBooktitle{Applied Networking Research Workshop (ANRW '21), July 24--30, 2021, Virtual Event, USA}
\acmPrice{15.00}
\acmDOI{10.1145/3472305.3472319}
\acmISBN{978-1-4503-8618-0/21/07}

\usepackage{hyperref}
\usepackage{units}

\usepackage{acro}
\makeatletter
\@ifpackagelater{acro}{2017/02/17}{\@ifpackagelater{acro}{2019/09/23}{}{
    \let\IDeclareAcronym\DeclareAcronym
    \renewcommand{\DeclareAcronym}[2]{%
        \IDeclareAcronym{#1}{%
        #2,foreign-plural={}
        }
    }
}}{}

\DeclareAcronym{EFM}{
  short        = EFM,
  long         = Explicit Flow Measurement
}

\DeclareAcronym{RTT}{
  short        = RTT,
  long         = round-trip time
}

\DeclareAcronym{L}{
  short        = L-Bit,
  long         = Loss Event Bit
}

\DeclareAcronym{Q}{
  short        = Q-Bit,
  long         = Square Bit
}

\DeclareAcronym{R}{
  short        = R-Bit,
  long         = Reflection Square Bit
}

\DeclareAcronym{T}{
  short        = T-Bit,
  long         = Round Trip Loss Bit
}

\DeclareAcronym{VEC}{
  short        = VEC,
  long         = Valid Edge Counter
}

\newcommand{\afblock}[1]{\noindent{\textbf{#1}}}
\newcommand{\sref}[1]{Sec.\,\ref{#1}}
\newcommand{\fig}[1]{Fig.\,\ref{#1}}
\newcommand{\ie}{i.e.,}
\newcommand{\eg}{e.g.,}

\begin{document}

\title{L, Q, R, and T - Which Spin Bit Cousin Is Here to Stay?}

\author{Ike Kunze, Klaus Wehrle, Jan R\"uth}
\email{{kunze, wehrle, rueth}@comsys.rwth-aachen.de}
\affiliation{%
  \institution{RWTH Aachen University}
}

\renewcommand{\shortauthors}{Kunze et al.}

\begin{abstract}
Network operators utilize traffic monitoring to locate and fix faults or performance bottlenecks.
This often relies on intrinsic protocol semantics, \eg{} sequence numbers, that many protocols share implicitly through their packet headers.
The arrival of (almost) fully encrypted transport protocols, such as QUIC, significantly complicates this monitoring as header data is no longer visible to passive observers.
Recognizing this challenge, QUIC offers explicit measurement semantics by exposing the spin bit to measure a flow's RTT.
Ongoing efforts in the IETF IPPM working group argue to expose further information and enable the passive quantification of packet loss.
This work implements and evaluates four currently proposed measurement techniques (\emph{L}-, \emph{Q}-, \emph{R}-, and \emph{T}-bit).
We find that all techniques generally provide accurate loss estimations, but that longer algorithmic intervals for \emph{Q} and \emph{R}, yet foremost for \emph{T}, complicate detecting very small loss rates or loss on short connections. 
Deployment combinations of \emph{Q} \& \emph{R} as well as \emph{Q} \& \emph{L}, thus, have the best potential for accurately grasping the loss in networks.  
\end{abstract}



%\keywords{Explicit flow measurements, loss measurements}

\maketitle

\section{Introduction} % (fold)
\label{sec:introduction}
The growing complexity of networks makes monitoring their performance increasingly challenging.
Yet, accurate information on the current network status is vital so that operators can act on errors in a timely manner or quickly identify misbehaving flows.
The types of available monitoring solutions significantly depend on the kind of observed network.
Datacenters, \eg{} allow for detailed monitoring~\cite{huang:sigcomm2020:omnimon} as operators can leverage both, passive measurements of existing production traffic and carefully injected active measurement traffic to obtain a complete picture of the network status.
In contrast, telco operators are mostly limited to passive monitoring as comprehensive active measurements are either infeasible or would not provide representative information~\cite{cociglio:AcmToN2019:multipoint}.

These passive measurements often rely on specific semantics of the deployed transport protocols.
For example, packet loss can be estimated using TCP's sequence numbers and acknowledgments~\cite{benko:globecom2002:tcppacketloss}.
The ongoing shift to (almost) fully encrypted transports~\cite{rueth:pam18:quicinthewild}, such as QUIC~\cite{RFC9000}, challenges these approaches as intrinsic protocol semantics are no longer visible to passive observers and thus lost for measurements.

Instead, new ways for enabling passive measurements are required.
Allman et al.~\cite{allman:2017:ccr:measurability} argue that measurability should be explicitly included into the protocols.
Adhering to this line of thought, QUIC incorporates a dedicated measurement bit, the spin bit, which is visible to passive observers and enables \ac{RTT} measurements~\cite{devaere:2018:imc:threebits}.

The spin bit's concept is similar to \emph{alternate marking} techniques, \eg{} defined in RFC 8321~\cite{RFC8321}, whose idea is to modulate signals onto production traffic by carefully controlling the values of designated measurement bits which an observer can then measure.
However, these concepts are deployed entirely in the network without cooperation or knowledge of the end-hosts.
Currently, the IETF IPPM working group~\cite{ietfippm:ietf:ippmwg} discusses ways to bring these principles into end-host protocols.
There are four distinct proposals for such \emph{\acp{EFM}}, called the L-, Q-, R-, and T-Bits~\cite{ietf-draft-explicit-flow-measurements} that enable packet loss measurements.
Yet, there is little well-documented knowledge on their effectiveness in measuring packet loss and next to no academic research.

In this paper, we thus evaluate the four proposals using Mininet~\cite{lantz2010network} and find that they mostly yield reasonable loss estimates.
Yet, longer algorithmic phases make some approaches miss very seldom loss or loss on short connections.
Specifically, this work contributes the following:
\begin{itemize}[noitemsep,topsep=0em,leftmargin=10pt]
  \item We implement the end-host logic of the four measurement mechanisms in aioquic~\cite{laine:2021:aioquic} (source code available at ~\cite{kunze:2021:aioquic-mod}).
  \item We devise a Mininet-based study to observe the four mechanisms in different network settings (setup available at ~\cite{kunze:2021:mininet}).
  \item We find that all mechanisms provide decent results subject to random loss; however, Q-, R-, and T-Bit are challenged in times of burst loss or short transmissions.
\end{itemize}

\afblock{Structure.}
\sref{sec:alternate_marking_schemes} introduces \acp{EFM}, discusses related work, and presents the four mechanisms under study.
We then discuss our Mininet-based testbed in \sref{sec:methodology} before evaluating the mechanisms in three scenarios.
First, we inspect their behavior subject to random loss (\sref{sub:random_loss}), and, second, to burst loss (\sref{sub:burst_loss}).
Lastly, we evaluate the impact of different flow lengths in \sref{sub:flow_lengths}.
Finally, \sref{sec:conclusion} concludes the paper.
% section introduction (end)

\section{Explicit Flow Measurements} % (fold)
\label{sec:alternate_marking_schemes}

Although having been an integral part of the early Internet~\cite{kleinrock:IEEECommunications2010:InternetHistory}, network measurements have mostly been designed independently from protocols and, as such, have usually relied on externally visible protocol semantics.
Prominent examples are measurements of the \ac{RTT} that base on the TCP handshake semantic~\cite{jiang:CCR2002:PassiveTCPMeasurements} or on the timestamp option~\cite{veal:pam2005:tcptimestamps,strowes:ACMQueue2013:tcptimestamps}.
However, as argued for and discussed by Allman et al.~\cite{allman:2017:ccr:measurability}, explicit measurement capabilities integrated into protocols are desirable for a number of reasons, \eg{} for providing a solid framework for making inferences on the network state.

\subsection{Related Work} % (fold)
\label{sub:related_work}
In this context, principles described in RFC 8321~\cite{RFC8321} have sparked a new push towards embedding measurements into protocols.
While the original \emph{alternate marking} is intended for measurements within the network, \ie{} without the participation of end-hosts, its principle of modeling measurable signaling information onto a communication channel has been picked up for the application to end-to-end flow measurements, also called \emph{\acfp{EFM}}.
QUIC even includes one instance following this scheme for enabling \ac{RTT} measurements: the spin bit~\cite{RFC9000}.
In essence, the server of a connection always \emph{reflects} the received spin bit, while the client \emph{flips} the spin bit after one RTT, thus creating an oscillating pattern.
On-path observers can collect the end-to-end RTT by measuring the time between changes of the spin bit value as the half period of the square wave signal produced by the spin bit equals the RTT.

De Vaere et al.~\cite{devaere:2018:imc:threebits} present and evaluate the spin bit.
In addition to the version incorporated in QUIC, they also describe a more sophisticated variant, the \emph{\ac{VEC}}, which requires two additional bits and allows for filtering out invalid spin bit edges, \eg{} due to reordering.
Using a Mininet-based evaluation and traffic passing through real networks, the authors show the general applicability of the spin bit.
Bulgarella et al.~\cite{bulgarella:2019:anrw:delaybit} propose a simplified version called the delay bit, which only requires one additional bit.
In Mininet experiments, they show that their variant achieves similar performance as the \ac{VEC}.
Meanwhile, the delay bit has also been proposed as a standalone technique~\cite{ietf-draft-explicit-flow-measurements}.

In this work, we focus on \ac{EFM}-based mechanisms to measure packet loss~\cite{ietf-draft-explicit-flow-measurements}: the L-, Q-, R-, and T-Bits, which we will explain in the following sections.
There is little work overall and especially no published academic work on the loss mechanisms.
At IETF, Ferrieux et al.~\cite{ferrieux:ietf105maprg:efm} provide an initial analysis of the Q- and \acsp{L}.
They present measurements between a CDN and an ISP and compare results obtained in different countries.
Additionally, they remark that flows have to have a certain length for the \acs{Q} to be applicable.

% subsection related_work (end)

\subsection{\ac{EFM}-based Loss Measurements} % (fold)
\label{sub:am_based_packet_loss_measurements_}

Cociglio et al.~\cite{ietf-draft-explicit-flow-measurements} currently propose four mechanisms for enabling packet loss measurements.
While the \acs{L} builds upon sender-side loss-detection provided by the transport protocol, the other three approaches rely on the periodic transmission of marked packets.
The \acs{Q} creates a constant square wave signal by flipping a bit after sending a fixed number of packets.
The \acs{R} builds upon this by itself marking packets for received Q bits.
The \acs{T} creates an initial train of packets that is then reflected several times between the server and the client.
In the following, we briefly outline the approaches, how they enable packet loss measurements, as well as end-host and observer logics; please refer to~\cite{ietf-draft-explicit-flow-measurements} for an in-depth explanation.
\sref{sub:discussion} then jointly discusses the techniques and analyzes their differences, especially regarding their loss observation capabilities.

\subsubsection{\acf{L}} % (fold)
\label{ssub:loss_event_bit_}

The \ac{L} relies on loss-detection by the transport protocol and essentially reports the number of packets that have been declared lost to the network.
For this, whenever the sender detects a loss, she increases a $counter_{L}$ by one.
To convey the number of lost packets to an on-path observer, she sets the \ac{L} (in the observable packet header) in the next packet and decreases the $counter_{L}$ by one and does so as long as it is above zero.
Thus, for each packet lost, the sender marks exactly one packet.

\afblock{End-Host Logic.}
The sender only has to check the current value of $counter_{L}$ and set the \ac{L} accordingly.
However, it requires a loss-detection mechanism on the transport layer.

\afblock{Observer Logic.}
An on-path observer can simply count the number of observed \ac{L} markings to determine the number of lost packets and might additionally count the number of overall transmitted packets to determine a loss rate.
% subsubsection loss_event_bit_ (end)

\subsubsection{\acf{T}} % (fold)
\label{ssub:round_trip_loss_bit_}

The \ac{T} is a multi-phase algorithm whose phases synchronize to the spin-bit periods.
In a first step, spanning at least two spin-bit periods, the client transmits a train of packets with a set \ac{T} (\emph{generation train}).
The server then reflects as many packets with a set \ac{T} as it has received, \ie{} lost packets are not reflected.
After a pause phase that spans at least one spin-bit period, the reflection phase begins.
In this phase, the client first transmits as many packets with a set \ac{T} to the server as it has received from the server in the previous reflection, \ie{} again marked packets may have been lost.
Finally, the server reflects all packets with a set \ac{T} that it has received.
At the end of this process, there is again a pause phase.

\afblock{End-Host Logic.}
The main task for hosts implementing the \ac{T} is to correctly assign received \ac{T} packets to the different algorithm phases and then set the lengths of outgoing phases accordingly.
The burden of deciding when to start which phase entirely lies on the client.
The main challenge is that the underlying spin-bit periods, and with them the envisioned state transitions, are known to be susceptible to reordering~\cite{devaere:2018:imc:threebits}, potentially causing faulty state transitions.

\afblock{Observer Logic.}
Observers can determine lost packets between the generation and reflection phase by comparing the number of marked packets in both phases.
For this, they have to detect and distinguish the different algorithm phases.
The pause phases simplify the detection, while generation and reflection can be distinguished by comparing observations of two subsequent phases and choosing the one with a greater value as the generation phase. 
% subsubsection round_trip_loss_bit_ (end)

\subsubsection{\acf{Q}} % (fold)
\label{ssub:square_bit_}
A \ac{Q} sender creates a square wave signal by first sending $N$ packets (a Q-Block) with an unset \ac{Q} and subsequently $N$ packets with a set \ac{Q}.

\afblock{End-Host Logic.}
The sender only has to alternate between the two signal levels after a fixed number of packets.
Our implementation uses a Q-Block length of $N=64$.

\afblock{Observer Logic.}
If the value of $N$ is known, on-path observers can simply compare the number of marked\-/\-un\-marked packets during a Q-Block with the expected number $N$ to determine the number of lost packets.
If $N$ is unknown, the observers can deduce $N$ from the number of received packets as Cociglio et al.~\cite{ietf-draft-explicit-flow-measurements} propose choosing $N$ as a power of 2.
In both cases, the observers have to correctly assign the counted packets to Q-Blocks even when reordering occurs near to the flanks of the square waves.
As a solution, Cociglio et al.~\cite{ietf-draft-explicit-flow-measurements} propose to add fixed thresholds to the phase change detection.
Our implementation uses a threshold of 8 so that a new signal level is only assumed after 8 values of the new phase have been detected, while all intermediate values are still assigned to the previous phase. 
% subsubsection square_bit_ (end)

\subsubsection{\acf{R}} % (fold)
\label{ssub:reflection_square_bit_}

The \ac{R} is co-designed with and builds upon the \ac{Q}, essentially using the same mechanism.
Initially, the \ac{R} is unset on outgoing packets.
After the first Q-Block has been received, the \ac{R} is toggled and its value is set on as many packets as have been received during the Q-Block.
It then flips again and similarly denotes the number of packets in the subsequently received Q-Block and so forth.
Since Q and R periods may overlap in time, \eg{} when the packet count in each direction is highly asymmetric%.}
, the number of packets that need R-marking is adjusted when one or more new Q-Blocks end before the current R-Block is completed by setting it to the average number of packets per Q-Block since the last R-Block started.
As such, the \ac{R} conveys the average number of packets received per Q-Block, thus enabling the observation of a statistical loss rate.

\afblock{End-Host Logic.}
The sender has to incorporate observer logic for the \ac{Q}.
More specifically, she has to to be aware of incoming \ac{Q} phases and detect Q-Block boundaries even in light of reordering to correctly count the packets belonging to each phase.
Additionally, she has to keep track of the average and properly react to changes such as the average dropping below the current R-marking count.

\afblock{Observer Logic.}
In addition to the consideration regarding reordering for the \ac{Q}, there is one significant challenge for an \ac{R} observer: she cannot know the number of initially transmitted \ac{R} packets.
Thus, she can only compare the counted \ac{R} packets to $N$, i.e., the initial Q-Block length.

This directly brings up the question of what conclusions can be drawn from the measurements of the different mechanisms.
In the following considerations, we first discuss the varying path resolutions of the different mechanisms before concluding our theoretical findings in a general discussion.
% subsubsection reflection_square_bit_ (end)

\subsection{Discussion} % (fold)
\label{sub:discussion}

While all of the four techniques enable passive packet loss measurements, they differ in the complexity of the end-host and observer logic, and in their path resolution, \ie{} which conclusions on the state of the network can be drawn from the measurements when seeing what portion of the traffic.

\subsubsection{Path Resolution} % (fold)
\label{ssub:path_resolution}
The accuracy and applicability of the techniques depends on where an observer monitors traffic and in which direction.
The following rationale is along the lines presented by Cociglio et al.~\cite{ietf-draft-explicit-flow-measurements}.
\fig{fig:testbed-path-resolution} visualizes a network consisting of one client, one server, and an observer in between and highlights which parts of the network are covered by each technique.
We first focus on a downstream observer that monitors traffic from the server to the client.

\afblock{\ac{Q}.}
The \ac{Q} allows for measuring packet loss on the link \emph{Downstream 1} as the observer can simply compare the counted number of \acp{Q} to the Q-Block length $N$.

\afblock{\ac{L}.}
The \ac{L} captures the overall loss of \emph{Downstream 1} + \emph{Downstream 2} (\emph{downstream loss}).
However, loss on \emph{Downstream 1} may skew the observations as \ac{L} markings can be lost prior to the observation point.

\afblock{\ac{R}.}
The \ac{R} resolution is even broader as a uni-di\-rec\-tion\-al observer can only determine the \emph{three-quarters loss}~\cite{ietf-draft-explicit-flow-measurements}, \ie{} the packet loss between the client and the down\-stream-side observer (\emph{Downstream 1}, \emph{Upstream 1}, and \emph{Upstream 2}).

\afblock{\ac{T}.}
Finally, the \ac{T} yields loss estimates for the observer-to-observer path, \ie{} the entire communication path.

Note that the same considerations apply for an observer that only monitors upstream traffic.
Further observations are possible when combining different approaches (the draft~\cite{ietf-draft-explicit-flow-measurements}, e.g., suggests Q+R-Bit and Q+L-Bit) or when considering a bi-directional observer.
Those are out of scope of this work.
% subsubsection path_resolution (end)

\subsubsection{General Discussion} % (fold)
\label{ssub:general_discussion}

Based on the presentation in this section, we can already make a few statements regarding the applicability of the different variants.
Multi-stage algorithms, such as the \ac{R} (as it depends on Q) and the \ac{T}, require more complex state handling and are thus more difficult to implement.
The \ac{Q} and \ac{L} are straightforward to implement, both on the end-hosts and on the observers, although mechanisms to handle reordering are required for the \ac{Q}.

The \ac{T} alone allows for the most flexible determination of packet loss across all network segments, especially if the observer is bi-directional.
According to the IETF IPPM mailing list, the combination of Q- and \acp{R}, as well as the \ac{T} variant are currently under testing at Telecom Italia~\cite{nilo:2021:ippm-mail}.
The Q+L combination is deployed at Akamai and Orange~\cite{lubashev:2021:ippm-mail} and also featured in the lsquic~\cite{litespeed:2021:lsquic} implementation.
This already indicates that the different combinations of the loss mechanisms seem to be deployable effectively.

After this general discussion, we next provide an experimental comparison of the different loss mechanisms for which we will first discuss our methodology.
% subsubsection general_discussion (end)
% subsection discussion (end)
% section alternate_marking_schemes (end)

\section{Methodology} % (fold)
\label{sec:methodology}
As discussed in the previous \sref{sec:alternate_marking_schemes}, the four mechanisms differ significantly in their fundamental working principles and in their path resolution which makes a comprehensive comparison challenging.
We consequently focus on evaluating the core characteristics of the mechanisms and do not investigate possible algorithm combinations or the possibility to locate loss on specific network segments.
In the following, we describe our evaluation setup in more depth. 

\begin{figure}[t]
  \center
  \includegraphics[width=\columnwidth]{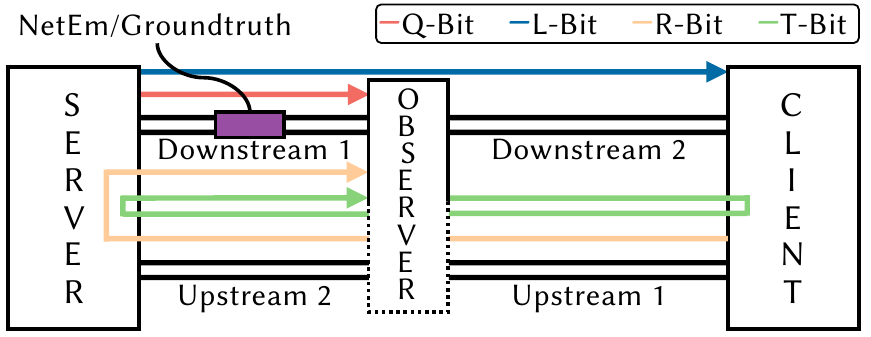}
  \caption{Our Mininet-based topology consists of two end-hosts (\emph{Server}, \emph{Client}) and one intermediate \emph{Observer}.
  A netem-based arbiter induces network impairments on \emph{Downstream 1}.}
  \label{fig:testbed-path-resolution}
  \vspace{-.5em}
\end{figure}

\afblock{Network Scenario.}
We perform our study in a Mininet-based testbed visualized in \fig{fig:testbed-path-resolution} which consists of two end-hosts (\emph{Server}, \emph{Client}) and one intermediate switch (\emph{Observer}).
Similar approaches have already been used for investigating the spin bit~\cite{devaere:2018:imc:threebits} and the delay bit~\cite{bulgarella:2019:anrw:delaybit}.
We shape a steady base delay of \unit[10]{ms} on all links using \emph{tc netem}.

\afblock{Network Impairments.}
To study the effects of different loss characteristics, we select link \emph{Downstream 1} for our experiments as a downstream observer can detect loss occurring on this link using all four mechanisms.
For inducing the impairments, we place a dedicated network arbiter on \emph{Downstream 1} which we then configure using \emph{tc netem}.

\afblock{EFM Implementation.}
For our experiments, we utilize the aioquic~\cite{laine:2021:aioquic} QUIC implementation.
We extend it by implementing all four loss mechanisms and additionally extend the short header by an additional byte (without header protection) that we insert behind the original spin bit.
This way, we can capture all mechanisms at the same time and compare their behavior subject to the same traffic.
Our modified aioquic implementation is available at \cite{kunze:2021:aioquic-mod}.

\afblock{Measurements.}
We do not perform online observations.
Instead, we capture incoming traffic on our \emph{Observer} using \emph{tcpdump} and then determine the behavior of the mechanisms offline.
We additionally derive a groundtruth of the actual loss as we can only configure loss probabilities and the true loss can thus vary.
For this, we collect queue statistics on our network arbiter using \emph{eBPF} which we then correlate with general statistics on the traffic coming into the arbiter which we again capture using \emph{tcpdump}.

In our analysis, we can derive the results obtained by the observer using our traffic captures and then compare them to groundtruth values based on our network arbiter.
% section methodology (end)

\section{Evaluation} % (fold)
\label{sec:evaluation}

Our evaluation consists of three parts.
First, we judge the general effectiveness of EFM-based loss measurements subject to idealized random loss behavior (\sref{sub:random_loss}).
In this setting, we generate symmetric traffic between \emph{Client} and \emph{Server} that is not congestion-controlled similar to \cite{bulgarella:2019:anrw:delaybit} and \cite{devaere:2018:imc:threebits}.
We then investigate the effect of burst losses while still keeping congestion control disabled (\sref{sub:burst_loss}).
Finally, we enable congestion control and switch to an asymmetric traffic scheme mirroring a typical download, \ie{} \emph{Client} requests files of different sizes from \emph{Server} (\sref{sub:flow_lengths}).
This enables investigating two aspects, 1) the performance of the \ac{EFM}-based schemes subject to fluctuations in the sending rates, and 2), the impact of connection durations on the expressiveness of the measurements, \ie{} whether they are equally effective for short-lived requests.

For each scenario, we perform 30 independent measurements and derive cumulative loss percentages as derived by the \ac{EFM} schemes, \ie{} loss rates summarizing a whole measurement run.
If not stated otherwise, we report the mean over all runs together with 99\% confidence intervals (CIs).

\subsection{Random Loss} % (fold)
\begin{figure}[t]
  \center
  \includegraphics[width=\columnwidth]{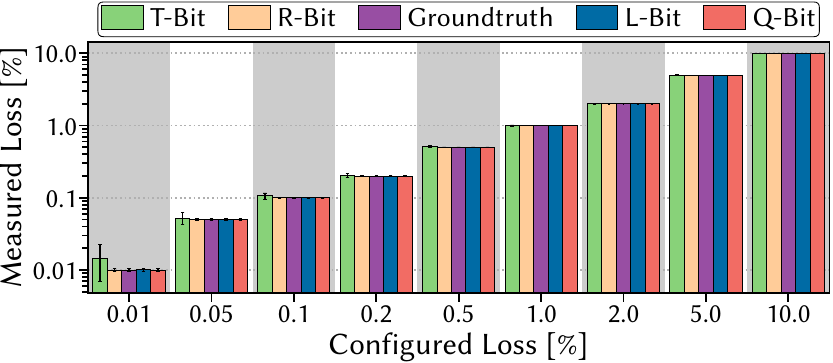}
  \caption{Mean loss rates with 99\% CIs reported by the four mechanisms as well as the groundtruth for different random loss rates.}
  \label{fig:baseline-lossrandom}
  \vspace{-.5em}
\end{figure}

\label{sub:random_loss}
The first network impairment in our study is random loss to investigate the general effectiveness of the \ac{EFM}-based loss mechanisms.
Using \textit{netem loss random}, we configure different random loss rates ranging from 0.01\% to 10\%.
To obtain statistically meaningful results, even for the very low loss percentages, we transmit approximately one million packets in each experiment run.
Following the methodology described by Biondini~\cite{biondini:elsevier2015:statisticshandbook}, this number is required to allow for an acceptable resolution of the groundtruth and should provide a relation of 0.1 between standard deviation and expected value of the resulting loss samples.
\fig{fig:baseline-lossrandom} shows the estimated loss rates as well as the groundtruth on a logarithmic scale for the different mechanisms across the different scenarios.

As can be seen, all measurement techniques are able to derive relatively accurate loss rates that are very close to the groundtruth values.
Only the \ac{T} has problems correctly determining the low loss percentages as its lengthy pause periods, accounting for at least one third of each measurement cycle, make it easily miss lost packets.

\begin{figure}[t]
  \center
  \includegraphics[width=\columnwidth]{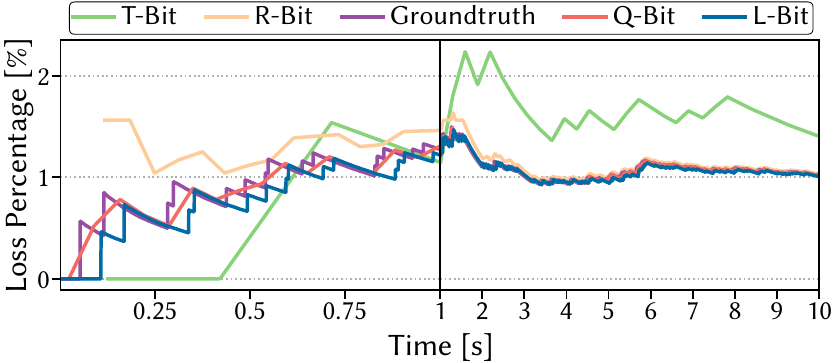}
  \caption{Cumulative loss rates of one selected setting with a configured mean random loss rate of 1\% within the 1st second (left) and over 9 more seconds (right).}
  \label{fig:baseline-lossrandom-time-course}
  \vspace{-.5em}
\end{figure}

\afblock{A closer look.}
To get a better feeling for the actual behavior of the different mechanisms, we next closely inspect the accuracy across one example run.
\fig{fig:baseline-lossrandom-time-course} illustrates the first second (left) and nine more seconds (right) of one hand-picked measurement run in the 1\% random loss setting.
Note that all data points represent the total loss as provided by the different mechanisms at the time of loss detection.

A first result is that the mechanisms provide different numbers of observations due to their specific algorithmic periodicities.
The \ac{Q}, \eg{} requires 64 transmitted packets before it can report a measurement.
The \ac{R} needs further 64 packets, as it first needs 64 in one direction that are then reflected.
Consequently, the first measurement of the \ac{R} comes after the first measurement of the \ac{Q}.
While the \ac{T} does not have fixed-length periods, it relies on the spin bit for its phase transitions which generally prolongs the interval between measurements and causes it to provide the fewest measurements.
We observe that the \ac{L} best tracks the evolution of the groundtruth, but as it relies on end-host loss detection, it slightly lags behind due to the end-hosts having to detect the loss in the first place.
Overall, we can see that the \ac{L} and the \ac{Q} are good approximations of the groundtruth while the \ac{R} and especially the \ac{T} require longer time to get close to the groundtruth.

While random loss enables deriving a baseline, loss often comes in bursts, \eg{} when network buffers overflow, causing a series of consecutive packets to be dropped.
Thus, we next investigate the impact of burst loss on the \ac{EFM} mechanisms.
% subsection random_loss (end)

\subsection{Burst Loss} % (fold)
\label{sub:burst_loss}
We model burst loss using the simple Gilbert model~\cite{gilbert:1960:burst-noise,elliot:1963:burst,hasslinger:2008:gilbert} and choose a fixed overall loss percentage of 1\% as this has shown adequate results in the random loss investigation.
We then derive corresponding parameter settings using the methodology described by Nasralla et al.~\cite{nasralla:temu2014:gemodelparams} to model different mean burst sizes.
Note that the corresponding state transition probabilities within the simple Gilbert model become very small when keeping the average loss rate at 1\%.
We thus again transmit approximately one million packets in each experiment to achieve statistically meaningful results.

\begin{figure}[t]
  \center
  \includegraphics[width=\columnwidth]{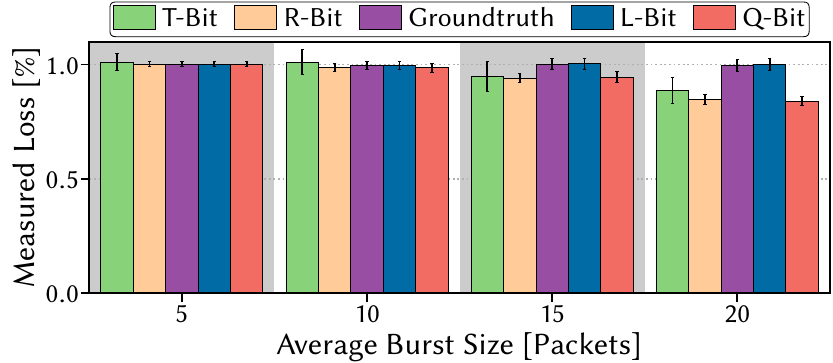}
  \caption{Mean loss rates with 99\% CIs reported by the four mechanisms and the groundtruth for different configured burst loss rates at a fixed loss rate of 1\%.}
  \label{fig:baseline-gemodel}
  \vspace{-.5em}
\end{figure}

\fig{fig:baseline-gemodel} depicts our results for four different average burst sizes.
As can be seen, the \ac{L} achieves a very high accuracy and is very close to the groundtruth for all investigated burst sizes.
In contrast, the other approaches show decreasing accuracy for larger burst sizes.
The main reason is that the loss detection techniques built into the transport protocols will eventually report on the lost packets in the case of the \ac{L}.
Thus, even though some L-marked packets will also get lost, the \ac{L} will eventually report on all lost packets.

In contrast, the other approaches rely entirely on signaling information and, furthermore, build upon periodic behavior.
In the worst case, entire periods might be wiped out if the burst sizes become too large.
For example, burst loss spanning an entire Q-Block length %, \ie{} 64 packets, 
will remain unnoticed by the observer.\footnote{The burst resilience can be extended to two Q-Blocks (cf.~\cite{ericsson:2021:spindump}), but this mechanism is not yet part of the EFM draft and not included here.}
Consequently, the \ac{R} provides similar results and, for both, increasing burst loss sizes cause an increasing difference between the real and the estimated loss rates.
The \ac{T}, on the other hand, shows a fluctuating behavior with larger confidence intervals.
This is due to the fact that the observer needs to resynchronize to the \ac{T} phases, thus yielding many invalid data points and thus reducing the overall number of measurements.

While the previous two settings are well-suited to investigate the general performance of the different \ac{EFM} mechanisms, real flows governed by congestion control would not keep steady transmission rates when loss appears as a proxy for congestion.
We thus next investigate the effect of enabling congestion control and of various flow lengths.
% subsection burst_loss (end)

\subsection{Flow Lengths} % (fold)
\label{sub:flow_lengths}
To study the usefulness of the \ac{EFM} mechanisms on more realistic traffic, we switch to the significantly asymmetric scenario of a download.
Additionally, we enable QUIC's congestion control (aioquic defaults to New Reno) to see whether this also plays a role. 
In contrast to the considerations in \sref{sub:random_loss} and \sref{sub:burst_loss}, we no longer ensure that there is enough traffic in each setting to achieve meaningful results, but instead explicitly aim to find out whether different flow lengths affect the accuracy of the loss mechanisms. 
We investigate the accuracy of the \ac{EFM} mechanisms at 1\% random loss with different download sizes, ranging from 50 kB, \ie{} around 40 packets, to 50 MB, \ie{} $40~000$ packets, so covering typical website sizes, streaming video chunks, or larger assets.

\begin{figure}[t]
  \center
  \includegraphics[width=\columnwidth]{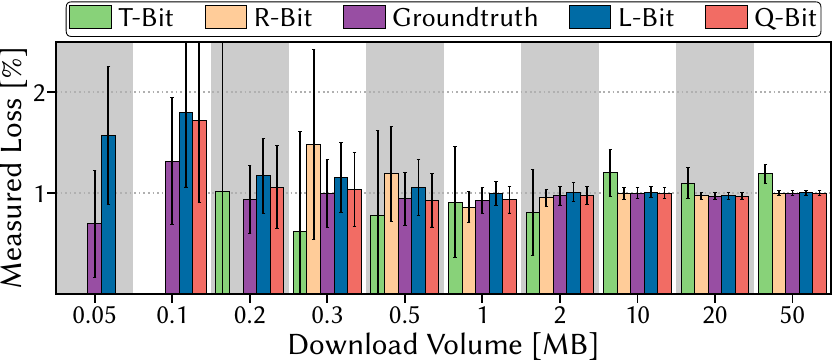}
  \caption{Mean loss rates with 99\% CIs reported by the four mechanisms and the groundtruth for different download volumes at a fixed random loss rate of 1\%.}
  \label{fig:congestion-lossrandom}
  \vspace{-.5em}
\end{figure}

For the small volume sizes, our results in \fig{fig:congestion-lossrandom} are very unstable, even the measured groundtruth varies a lot around the configured loss due to the small number of packets.
Another observation, that follows from \fig{fig:baseline-lossrandom-time-course}, is that the Q-, R- and \acp{T} require a certain amount of transmitted data to actually provide measurements, namely, in our case, 100kB (\ac{Q}), 200kB (\ac{T}), and 300kB (\ac{R}).
In contrast, the \ac{L} provides measurements right from the start.

Overall, we observe that the accuracy only reaches stable levels in the MB range.
The phase-based algorithms require longer flows before producing reliable results, and, as such, they appear less suited to capture loss of short-sized interactions.
Additionally, since \ac{T} and \ac{R} rely on receiver-generated packets, which, in this setting, are created through ACKs, the chosen ACK ratio further affects the applicability.
% subsection flow_lengths (end)
% section evaluation (end)

\section{Conclusion} % (fold)
\label{sec:conclusion}

\aclp{EFM} provide a novel way of explicitly monitoring connection statistics by passive observers in the network.
While measurements of the network delay are already incorporated in QUIC in the form of the spin bit, there is ongoing debate whether and how to enable packet loss measurements.
In this work, we investigate four proposals, L-, Q-, R-, and T-Bit, currently discussed in the IETF IPPM working group.
Using a Mininet testbed, we study the proposals across several network settings and find that they are generally capable of providing accurate measurements when subject to random loss.
This changes in times of burst loss, as longer algorithmic phases of Q-, R-, and T-Bit make them easily miss loss if an entire phase is wiped out.
On the other hand, the \ac{L} provides an accuracy close to the groundtruth, but requires an accurate loss detection by the transport.
Concluding, we find that the standalone \ac{L} or combinations, such as Q+\ac{R} or Q+\ac{L}, look the most promising, at least when only considering measurement accuracy.
While the required path resolution might also affect real deployment decisions, we leave corresponding investigations and considerations for future work.
% section conclusion (end)

%\begin{acks}
%\end{acks}

\clearpage
\balance
\bibliographystyle{ACM-Reference-Format}
\bibliography{literature}


\begin{thebibliography}{29}

%%% the Web address.

\ifx \showCODEN    \undefined \def \showCODEN     #1{\unskip}     \fi
\ifx \showDOI      \undefined \def \showDOI       #1{#1}\fi
\ifx \showISBNx    \undefined \def \showISBNx     #1{\unskip}     \fi
\ifx \showISBNxiii \undefined \def \showISBNxiii  #1{\unskip}     \fi
\ifx \showISSN     \undefined \def \showISSN      #1{\unskip}     \fi
\ifx \showLCCN     \undefined \def \showLCCN      #1{\unskip}     \fi
\ifx \shownote     \undefined \def \shownote      #1{#1}          \fi
\ifx \showarticletitle \undefined \def \showarticletitle #1{#1}   \fi
\ifx \showURL      \undefined \def \showURL       {\relax}        \fi
% invisible to TeX
\providecommand\bibfield[2]{#2}
\providecommand\bibinfo[2]{#2}
\providecommand\natexlab[1]{#1}
\providecommand\showeprint[2][]{arXiv:#2}

\bibitem[\protect\citeauthoryear{Allman, Beverly, and Trammell}{Allman
  et~al\mbox{.}}{2017}]%
        {allman:2017:ccr:measurability}
\bibfield{author}{\bibinfo{person}{Mark Allman}, \bibinfo{person}{Robert
  Beverly}, {and} \bibinfo{person}{Brian Trammell}.}
  \bibinfo{year}{2017}\natexlab{}.
\newblock \showarticletitle{{Principles for Measurability in Protocol Design}}.
\newblock \bibinfo{journal}{\emph{ACM SIGCOMM Computer Communication Review
  (CCR '17)}} \bibinfo{volume}{47}, \bibinfo{number}{2} (\bibinfo{year}{2017}).
\newblock
\urldef\tempurl%
\url{https://doi.org/10.1145/3089262.3089264}
\showDOI{\tempurl}


\bibitem[\protect\citeauthoryear{Benko and Veres}{Benko and Veres}{2002}]%
        {benko:globecom2002:tcppacketloss}
\bibfield{author}{\bibinfo{person}{Peter Benko} {and} \bibinfo{person}{Andras
  Veres}.} \bibinfo{year}{2002}\natexlab{}.
\newblock \showarticletitle{{A Passive Method for Estimating End-to-End TCP
  Packet Loss}}. In \bibinfo{booktitle}{\emph{IEEE Global Telecommunications
  Conference}} \emph{(\bibinfo{series}{GLOBECOM '02})}.
\newblock
\urldef\tempurl%
\url{https://doi.org/10.1109/GLOCOM.2002.1189102}
\showDOI{\tempurl}


\bibitem[\protect\citeauthoryear{Biondini}{Biondini}{2015}]%
        {biondini:elsevier2015:statisticshandbook}
\bibfield{author}{\bibinfo{person}{Gino Biondini}.}
  \bibinfo{year}{2015}\natexlab{}.
\newblock \showarticletitle{{An Introduction to Rare Event Simulation and
  Importance Sampling}}.
\newblock In \bibinfo{booktitle}{\emph{Handbook of Statistics}}.
  Vol.~\bibinfo{volume}{33}. \bibinfo{publisher}{Elsevier},
  \bibinfo{pages}{29--68}.
\newblock
\urldef\tempurl%
\url{https://doi.org/10.1016/B978-0-444-63492-4.00002-2}
\showDOI{\tempurl}


\bibitem[\protect\citeauthoryear{Bulgarella, Cociglio, Fioccola, Marchetto, and
  Sisto}{Bulgarella et~al\mbox{.}}{2019}]%
        {bulgarella:2019:anrw:delaybit}
\bibfield{author}{\bibinfo{person}{Fabio Bulgarella}, \bibinfo{person}{Mauro
  Cociglio}, \bibinfo{person}{Giuseppe Fioccola}, \bibinfo{person}{Guido
  Marchetto}, {and} \bibinfo{person}{Riccardo Sisto}.}
  \bibinfo{year}{2019}\natexlab{}.
\newblock \showarticletitle{{Performance Measurements of QUIC Communications}}.
  In \bibinfo{booktitle}{\emph{ACM Applied Networking Research Workshop}}
  \emph{(\bibinfo{series}{ANRW '19})}.
\newblock
\urldef\tempurl%
\url{https://doi.org/10.1145/3340301.3341127}
\showDOI{\tempurl}


\bibitem[\protect\citeauthoryear{Cociglio, Ferrieux, Fiocolla, Lubashev,
  Bulgarella, Hamchaoui, Nilo, Sisto, and Tikhonov}{Cociglio
  et~al\mbox{.}}{2021}]%
        {ietf-draft-explicit-flow-measurements}
\bibfield{author}{\bibinfo{person}{M. Cociglio}, \bibinfo{person}{A. Ferrieux},
  \bibinfo{person}{G. Fiocolla}, \bibinfo{person}{I. Lubashev},
  \bibinfo{person}{F. Bulgarella}, \bibinfo{person}{I. Hamchaoui},
  \bibinfo{person}{M. Nilo}, \bibinfo{person}{R. Sisto}, {and}
  \bibinfo{person}{D. Tikhonov}.} \bibinfo{year}{2021}\natexlab{}.
\newblock \bibinfo{booktitle}{\emph{{Explicit Flow Measurements Techniques}}}.
\newblock \bibinfo{type}{Internet-Draft}. \bibinfo{institution}{IETF}.
\newblock
\urldef\tempurl%
\url{https://datatracker.ietf.org/doc/draft-mdt-ippm-explicit-flow-measurements/}
\showURL{%
\tempurl}
\newblock
\shownote{Work in Progress.}


\bibitem[\protect\citeauthoryear{Cociglio, Fioccola, Marchetto, Sapio, and
  Sisto}{Cociglio et~al\mbox{.}}{2019}]%
        {cociglio:AcmToN2019:multipoint}
\bibfield{author}{\bibinfo{person}{Mauro Cociglio}, \bibinfo{person}{Giuseppe
  Fioccola}, \bibinfo{person}{Guido Marchetto}, \bibinfo{person}{Amedeo Sapio},
  {and} \bibinfo{person}{Riccardo Sisto}.} \bibinfo{year}{2019}\natexlab{}.
\newblock \showarticletitle{{Multipoint Passive Monitoring in Packet
  Networks}}.
\newblock \bibinfo{journal}{\emph{IEEE/ACM Transactions on Networking (TON
  '19)}} \bibinfo{volume}{27}, \bibinfo{number}{6} (\bibinfo{year}{2019}).
\newblock
\urldef\tempurl%
\url{https://doi.org/10.1109/TNET.2019.2950157}
\showDOI{\tempurl}


\bibitem[\protect\citeauthoryear{De~Vaere, B{\"u}hler, K{\"u}hlewind, and
  Trammell}{De~Vaere et~al\mbox{.}}{2018}]%
        {devaere:2018:imc:threebits}
\bibfield{author}{\bibinfo{person}{Piet De~Vaere}, \bibinfo{person}{Tobias
  B{\"u}hler}, \bibinfo{person}{Mirja K{\"u}hlewind}, {and}
  \bibinfo{person}{Brian Trammell}.} \bibinfo{year}{2018}\natexlab{}.
\newblock \showarticletitle{{Three Bits Suffice: Explicit Support for Passive
  Measurement of Internet Latency in QUIC and TCP}}. In
  \bibinfo{booktitle}{\emph{ACM Internet Measurement Conference}}
  \emph{(\bibinfo{series}{IMC '18})}.
\newblock
\urldef\tempurl%
\url{https://doi.org/10.1145/3278532.3278535}
\showDOI{\tempurl}


\bibitem[\protect\citeauthoryear{Elliott}{Elliott}{1963}]%
        {elliot:1963:burst}
\bibfield{author}{\bibinfo{person}{E.~O. Elliott}.}
  \bibinfo{year}{1963}\natexlab{}.
\newblock \showarticletitle{{Estimates of Error Rates for Codes on Burst-Noise
  Channels}}.
\newblock \bibinfo{journal}{\emph{Bell System Technical Journal}}
  \bibinfo{volume}{42}, \bibinfo{number}{5} (\bibinfo{year}{1963}),
  \bibinfo{pages}{1977--1997}.
\newblock
\urldef\tempurl%
\url{https://doi.org/10.1002/j.1538-7305.1963.tb00955.x}
\showDOI{\tempurl}


\bibitem[\protect\citeauthoryear{Ericsson}{Ericsson}{2021}]%
        {ericsson:2021:spindump}
\bibfield{author}{\bibinfo{person}{Ericsson}.} \bibinfo{year}{2021}\natexlab{}.
\newblock \bibinfo{title}{{Spindump on GitHub}}.
\newblock
\newblock
\urldef\tempurl%
\url{https://github.com/EricssonResearch/spindump}
\showURL{%
\tempurl}


\bibitem[\protect\citeauthoryear{Ferrieux, Hamchaoui, and Lubashev}{Ferrieux
  et~al\mbox{.}}{2019}]%
        {ferrieux:ietf105maprg:efm}
\bibfield{author}{\bibinfo{person}{Alexandre Ferrieux},
  \bibinfo{person}{Isabelle Hamchaoui}, {and} \bibinfo{person}{Igor Lubashev}.}
  \bibinfo{year}{2019}\natexlab{}.
\newblock \bibinfo{title}{{Packet Loss Signaling for Encrypted Protocols}}.
\newblock
  \bibinfo{howpublished}{\url{https://datatracker.ietf.org/meeting/105/materials/slides-105-maprg-packet-loss-signaling-for-encrypted-protocols-01.pdf}}.
\newblock


\bibitem[\protect\citeauthoryear{Fioccola, Capello, Cociglio, Castaldelli,
  Chen, Zheng, Mirsky, and Mizrahi}{Fioccola et~al\mbox{.}}{2018}]%
        {RFC8321}
\bibfield{author}{\bibinfo{person}{G. Fioccola}, \bibinfo{person}{A. Capello},
  \bibinfo{person}{M. Cociglio}, \bibinfo{person}{L. Castaldelli},
  \bibinfo{person}{M. Chen}, \bibinfo{person}{L. Zheng}, \bibinfo{person}{G.
  Mirsky}, {and} \bibinfo{person}{T. Mizrahi}.}
  \bibinfo{year}{2018}\natexlab{}.
\newblock \bibinfo{booktitle}{\emph{{Alternate-Marking Method for Passive and
  Hybrid Performance Monitoring}}}.
\newblock \bibinfo{type}{RFC} 8321. \bibinfo{institution}{IETF}.
\newblock
\urldef\tempurl%
\url{http://tools.ietf.org/rfc/rfc8321.txt}
\showURL{%
\tempurl}


\bibitem[\protect\citeauthoryear{Gilbert}{Gilbert}{1960}]%
        {gilbert:1960:burst-noise}
\bibfield{author}{\bibinfo{person}{E.~N. Gilbert}.}
  \bibinfo{year}{1960}\natexlab{}.
\newblock \showarticletitle{{Capacity of a Burst-Noise Channel}}.
\newblock \bibinfo{journal}{\emph{Bell System Technical Journal}}
  \bibinfo{volume}{39}, \bibinfo{number}{5} (\bibinfo{year}{1960}),
  \bibinfo{pages}{1253--1265}.
\newblock
\urldef\tempurl%
\url{https://doi.org/10.1002/j.1538-7305.1960.tb03959.x}
\showDOI{\tempurl}


\bibitem[\protect\citeauthoryear{Ha{\ss}linger and Hohlfeld}{Ha{\ss}linger and
  Hohlfeld}{2008}]%
        {hasslinger:2008:gilbert}
\bibfield{author}{\bibinfo{person}{Gerhard Ha{\ss}linger} {and}
  \bibinfo{person}{Oliver Hohlfeld}.} \bibinfo{year}{2008}\natexlab{}.
\newblock \showarticletitle{{The Gilbert-Elliott Model for Packet Loss in Real
  Time Services on the Internet}}. In \bibinfo{booktitle}{\emph{VDE GI/ITG
  Conference-Measurement, Modelling and Evalutation of Computer and
  Communication Systems}} \emph{(\bibinfo{series}{MMB '08})}.
\newblock
\urldef\tempurl%
\url{https://www.vde-verlag.de/proceedings-en/563090019.html}
\showURL{%
\tempurl}


\bibitem[\protect\citeauthoryear{Huang, Sun, Lee, Bai, Zhu, and Bao}{Huang
  et~al\mbox{.}}{2020}]%
        {huang:sigcomm2020:omnimon}
\bibfield{author}{\bibinfo{person}{Qun Huang}, \bibinfo{person}{Haifeng Sun},
  \bibinfo{person}{Patrick~PC Lee}, \bibinfo{person}{Wei Bai},
  \bibinfo{person}{Feng Zhu}, {and} \bibinfo{person}{Yungang Bao}.}
  \bibinfo{year}{2020}\natexlab{}.
\newblock \showarticletitle{{OmniMon: Re-architecting Network Telemetry with
  Resource Efficiency and Full Accuracy}}. In \bibinfo{booktitle}{\emph{ACM
  Special Interest Group on Data Communication on the Applications,
  Technologies, Architectures, and Protocols for Computer Communication}}
  \emph{(\bibinfo{series}{SIGCOMM '20})}.
\newblock
\urldef\tempurl%
\url{https://doi.org/10.1145/3387514.3405877}
\showDOI{\tempurl}


\bibitem[\protect\citeauthoryear{{IP Performance Measurement Working
  Group}}{{IP Performance Measurement Working Group}}{2021}]%
        {ietfippm:ietf:ippmwg}
\bibfield{author}{\bibinfo{person}{{IP Performance Measurement Working
  Group}}.} \bibinfo{year}{2021}\natexlab{}.
\newblock \bibinfo{title}{{IP Performance Measurement (ippm)}}.
\newblock
  \bibinfo{howpublished}{\url{https://datatracker.ietf.org/wg/ippm/about/}}.
\newblock


\bibitem[\protect\citeauthoryear{Iyengar and Thomson}{Iyengar and
  Thomson}{2021}]%
        {RFC9000}
\bibfield{author}{\bibinfo{person}{Jana Iyengar} {and} \bibinfo{person}{Martin
  Thomson}.} \bibinfo{year}{2021}\natexlab{}.
\newblock \bibinfo{booktitle}{\emph{{QUIC: A UDP-Based Multiplexed and Secure
  Transport}}}.
\newblock \bibinfo{type}{RFC} 9000. \bibinfo{institution}{Internet Engineering
  Task Force}.
\newblock
\urldef\tempurl%
\url{http://tools.ietf.org/rfc/rfc9000.txt}
\showURL{%
\tempurl}


\bibitem[\protect\citeauthoryear{Jiang and Dovrolis}{Jiang and
  Dovrolis}{2002}]%
        {jiang:CCR2002:PassiveTCPMeasurements}
\bibfield{author}{\bibinfo{person}{Hao Jiang} {and}
  \bibinfo{person}{Constantinos Dovrolis}.} \bibinfo{year}{2002}\natexlab{}.
\newblock \showarticletitle{{Passive Estimation of TCP Round-Trip Times}}.
\newblock \bibinfo{journal}{\emph{ACM SIGCOMM Computer Communication Review
  (CCR '02)}} \bibinfo{volume}{32}, \bibinfo{number}{3} (\bibinfo{year}{2002}).
\newblock
\urldef\tempurl%
\url{https://doi.org/10.1145/571697.571725}
\showDOI{\tempurl}


\bibitem[\protect\citeauthoryear{Kleinrock}{Kleinrock}{2010}]%
        {kleinrock:IEEECommunications2010:InternetHistory}
\bibfield{author}{\bibinfo{person}{Leonard Kleinrock}.}
  \bibinfo{year}{2010}\natexlab{}.
\newblock \showarticletitle{{An Early History of the Internet}}.
\newblock \bibinfo{journal}{\emph{IEEE Communications Magazine}}
  \bibinfo{volume}{48}, \bibinfo{number}{8} (\bibinfo{year}{2010}).
\newblock
\urldef\tempurl%
\url{https://doi.org/10.1109/MCOM.2010.5534584}
\showDOI{\tempurl}


\bibitem[\protect\citeauthoryear{Kunze~et al.}{Kunze~et al.}{2021a}]%
        {kunze:2021:aioquic-mod}
\bibfield{author}{\bibinfo{person}{I. Kunze~et al.}}
  \bibinfo{year}{2021}\natexlab{a}.
\newblock \bibinfo{title}{{aioquic on GitHub}}.
\newblock
\newblock
\urldef\tempurl%
\url{https://github.com/COMSYS/aioquic}
\showURL{%
\tempurl}


\bibitem[\protect\citeauthoryear{Kunze~et al.}{Kunze~et al.}{2021b}]%
        {kunze:2021:mininet}
\bibfield{author}{\bibinfo{person}{I. Kunze~et al.}}
  \bibinfo{year}{2021}\natexlab{b}.
\newblock \bibinfo{title}{{EFM evaluation framework on GitHub}}.
\newblock
\newblock
\urldef\tempurl%
\url{https://github.com/COMSYS/efm-evaluation-anrw}
\showURL{%
\tempurl}


\bibitem[\protect\citeauthoryear{{Lain{\'e}, Jeremy}}{{Lain{\'e},
  Jeremy}}{2021}]%
        {laine:2021:aioquic}
\bibfield{author}{\bibinfo{person}{{Lain{\'e}, Jeremy}}.}
  \bibinfo{year}{2021}\natexlab{}.
\newblock \bibinfo{title}{{aioquic on GitHub}}.
\newblock
\newblock
\urldef\tempurl%
\url{https://github.com/aiortc/aioquic}
\showURL{%
\tempurl}


\bibitem[\protect\citeauthoryear{Lantz, Heller, and McKeown}{Lantz
  et~al\mbox{.}}{2010}]%
        {lantz2010network}
\bibfield{author}{\bibinfo{person}{Bob Lantz}, \bibinfo{person}{Brandon
  Heller}, {and} \bibinfo{person}{Nick McKeown}.}
  \bibinfo{year}{2010}\natexlab{}.
\newblock \showarticletitle{{A Network in a Laptop: Rapid Prototyping for
  Software-Defined Networks}}. In \bibinfo{booktitle}{\emph{Proceedings of the
  9th ACM SIGCOMM Workshop on Hot Topics in Networks}}
  \emph{(\bibinfo{series}{Hotnets-IX})}.
\newblock
\showISBNx{9781450304092}
\urldef\tempurl%
\url{https://doi.org/10.1145/1868447.1868466}
\showDOI{\tempurl}


\bibitem[\protect\citeauthoryear{{LiteSpeed Technologies Inc}}{{LiteSpeed
  Technologies Inc}}{2021}]%
        {litespeed:2021:lsquic}
\bibfield{author}{\bibinfo{person}{{LiteSpeed Technologies Inc}}.}
  \bibinfo{year}{2021}\natexlab{}.
\newblock \bibinfo{title}{{lsquic on GitHub}}.
\newblock
\newblock
\urldef\tempurl%
\url{https://github.com/litespeedtech/lsquic}
\showURL{%
\tempurl}


\bibitem[\protect\citeauthoryear{Lubashev}{Lubashev}{2021}]%
        {lubashev:2021:ippm-mail}
\bibfield{author}{\bibinfo{person}{Igor Lubashev}.}
  \bibinfo{year}{2021}\natexlab{}.
\newblock \bibinfo{title}{Re: [ippm] Comparing Alternate Marking and Explicit
  Flow Measurements (Spin bit, ...)}.
\newblock \bibinfo{howpublished}{IETF IPPM Mailing List}.
\newblock
\urldef\tempurl%
\url{https://mailarchive.ietf.org/arch/msg/ippm/v1je6osxrRS1zBCruPqEBnDyw8Y/}
\showURL{%
\tempurl}


\bibitem[\protect\citeauthoryear{Nasralla, Hewage, and Martini}{Nasralla
  et~al\mbox{.}}{2014}]%
        {nasralla:temu2014:gemodelparams}
\bibfield{author}{\bibinfo{person}{Moustafa~M. Nasralla},
  \bibinfo{person}{Chaminda~T.E.R. Hewage}, {and} \bibinfo{person}{Maria~G.
  Martini}.} \bibinfo{year}{2014}\natexlab{}.
\newblock \showarticletitle{{Subjective and Objective Evaluation and Packet
  Loss Modeling for 3D Video Transmission over LTE Networks}}. In
  \bibinfo{booktitle}{\emph{IEEE International Conference on Telecommunications
  and Multimedia}} \emph{(\bibinfo{series}{TEMU '14})}.
\newblock
\urldef\tempurl%
\url{https://doi.org/10.1109/TEMU.2014.6917770}
\showDOI{\tempurl}


\bibitem[\protect\citeauthoryear{Nilo}{Nilo}{2021}]%
        {nilo:2021:ippm-mail}
\bibfield{author}{\bibinfo{person}{Massimo Nilo}.}
  \bibinfo{year}{2021}\natexlab{}.
\newblock \bibinfo{title}{[ippm] Explicit Flow Measurements implementations}.
\newblock \bibinfo{howpublished}{IETF IPPM Mailing List}.
\newblock
\urldef\tempurl%
\url{https://mailarchive.ietf.org/arch/msg/ippm/3EpGbp78CtimKTrEa4oKnDsXEZk/}
\showURL{%
\tempurl}


\bibitem[\protect\citeauthoryear{R{\"u}th, Poese, Dietzel, and
  Hohlfeld}{R{\"u}th et~al\mbox{.}}{2018}]%
        {rueth:pam18:quicinthewild}
\bibfield{author}{\bibinfo{person}{Jan R{\"u}th}, \bibinfo{person}{Ingmar
  Poese}, \bibinfo{person}{Christoph Dietzel}, {and} \bibinfo{person}{Oliver
  Hohlfeld}.} \bibinfo{year}{2018}\natexlab{}.
\newblock \showarticletitle{{A First Look at QUIC in the Wild}}. In
  \bibinfo{booktitle}{\emph{Springer Passive and Active Measurement}}
  \emph{(\bibinfo{series}{PAM '18})}.
\newblock
\urldef\tempurl%
\url{https://doi.org/10.1007/978-3-319-76481-8_19}
\showDOI{\tempurl}


\bibitem[\protect\citeauthoryear{Strowes}{Strowes}{2013}]%
        {strowes:ACMQueue2013:tcptimestamps}
\bibfield{author}{\bibinfo{person}{Stephen~D. Strowes}.}
  \bibinfo{year}{2013}\natexlab{}.
\newblock \showarticletitle{{Passively Measuring TCP Round-Trip Times}}.
\newblock \bibinfo{journal}{\emph{Communications of the ACM (CACM '13)}}
  \bibinfo{volume}{56}, \bibinfo{number}{10} (\bibinfo{date}{Oct.}
  \bibinfo{year}{2013}), \bibinfo{pages}{57--64}.
\newblock
\urldef\tempurl%
\url{https://doi.org/10.1145/2507771.2507781}
\showDOI{\tempurl}


\bibitem[\protect\citeauthoryear{Veal, Li, and Lowenthal}{Veal
  et~al\mbox{.}}{2005}]%
        {veal:pam2005:tcptimestamps}
\bibfield{author}{\bibinfo{person}{Bryan Veal}, \bibinfo{person}{Kang Li},
  {and} \bibinfo{person}{David Lowenthal}.} \bibinfo{year}{2005}\natexlab{}.
\newblock \showarticletitle{{New Methods for Passive Estimation of TCP
  Round-Trip Times}}. In \bibinfo{booktitle}{\emph{Spinger Passive and Active
  Measurement Conference}} \emph{(\bibinfo{series}{PAM '05})}.
\newblock
\urldef\tempurl%
\url{https://doi.org/10.1007/978-3-540-31966-5_10}
\showDOI{\tempurl}


\end{thebibliography}

\end{document}